\documentclass[aps,prd,preprint,superscriptaddress,tightenlines,nofootinbib]{revtex4}

\usepackage{graphicx}
\usepackage{dcolumn}
\usepackage{bm}

\begin{document}

\preprint{CLNS 03/1813}       
\preprint{CLEO 03-01}         

\title{Measurements of the branching fractions and helicity amplitudes 
       in $B \to D^* \rho$ decays}

\author{S.~E.~Csorna}
\author{I.~Danko}
\affiliation{Vanderbilt University, Nashville, Tennessee 37235}
\author{G.~Bonvicini}
\author{D.~Cinabro}
\author{M.~Dubrovin}
\author{S.~McGee}
\affiliation{Wayne State University, Detroit, Michigan 48202}
\author{A.~Bornheim}
\author{E.~Lipeles}
\author{S.~P.~Pappas}
\author{A.~Shapiro}
\author{W.~M.~Sun}
\author{A.~J.~Weinstein}
\affiliation{California Institute of Technology, Pasadena, California 91125}
% ...
\collaboration{CLEO Collaboration} 
\noaffiliation


\date{January 15, 2003}

\begin{abstract}
\noindent Using $9.1$ fb$^{-1}$ of $e^+ e^-$ data collected at the 
$\Upsilon(4S)$ with the CLEO detector using 
the Cornell Electron Storage Ring, measurements are 
reported for both the branching fractions and the helicity amplitudes for 
the decays $B^-\to D^{*0}\rho^-$ and $\bar{B}^0\to D^{*+}\rho^-$. 
The fraction of longitudinal polarization in $\bar{B}^0\to D^{*+}\rho^-$ 
is found to be consistent with that in $\bar{B}^0\to D^{*+}\ell^-\bar{\nu}$ 
at $q^2=M^2_{\rho}$, 
indicating that the factorization approximation works well.  
The longitudinal polarization in the $B^-$ mode is similar.  
The measurements also show evidence of non-trivial final-state interaction 
phases for the helicity amplitudes.  
\end{abstract}

\maketitle


Hadronic decays of heavy mesons are complicated by final-state interactions 
(FSI) which result from strong re-scattering of the products of the 
weak decay process.  
FSI effects may be less important if the final state is easy to produce 
directly via weak decay.  
It is also argued that if the final-state hadrons 
separate rapidly, due to a large energy release, there is little 
time for interaction.  
The factorization hypothesis, widely used in heavy-quark physics 
for hadronic two-body decays~\cite{ZPC34_103}, 
assumes that the two hadronic currents may be treated independently 
of each other, neglecting FSI.  
In particular, the BSW model~\cite{ZPC34_103} utilizes this approximation 
in assuming that the short and long distance QCD contributions can be 
factorized.  
However, the validity of the factorization hypothesis 
has not been demonstrated by any rigorous theoretical calculation. 

K\"{o}rner and Goldstein~\cite{PL89B_105} suggest a test 
of the factorization hypothesis by examining the polarization in 
$B$ meson decays into two vector mesons.  The idea is that, under 
the factorization hypothesis, certain hadronic decays are analogous 
to similar semileptonic decays evaluated at a fixed value 
of the momentum transfer, $q^2 \equiv M_{\ell\bar{\nu}}^2$.  
For instance, 
the polarization of the $D^{*+}$ meson in $\bar{B}^0 \to D^{*+}\rho^-$ 
should equal that in $\bar{B}^0 \to D^{*+}l^-\bar{\nu}$ at $q^2=M_{\rho}^2$.  
Specifically: 
\begin{equation}
  \displaystyle\frac{\Gamma_L}{\Gamma}(\bar{B}^0\to D^{*+}\rho^-) = \left.
  \displaystyle\frac{\Gamma_L}{\Gamma}(\bar{B}^0\to D^{*+}\ell^-\bar{\nu})
  \right|_{q^2=M_{\rho}^2}.
\end{equation}
Here, $\Gamma_L/\Gamma$ is the fraction of longitudinal polarization.

The differential decay rate for $B \to D^* \rho^-$ can be expressed 
in terms of three complex helicity amplitudes $H_0$, $H_+$ and $H_-$ as: 
\begin{equation}
\begin{array}{l}
\displaystyle\frac{d^3\Gamma}{d\cos\theta_{D^*}d\cos\theta_{\rho}d\chi}  = 
\displaystyle\frac{9}{32\pi}\times
\Big\lbrace4|H_0|^2\cos^2\theta_{D^*}\cos^2\theta_{\rho} +
\left(|H_+|^2 + |H_-|^2\right)\sin^2\theta_{D^*}\sin^2\theta_{\rho}\\[4mm]
\quad\quad\quad\quad + 2 \left[\Re (H_+H^*_-)\cos 2\chi - 
\Im (H_+H^*_-)\sin 2\chi\right]\sin^2\theta_{D^*}\sin^2\theta_{\rho} \\[2mm]
\quad\quad\quad\quad + \left[\Re (H_+H^*_0 + H_-H^*_0)\cos\chi - 
\Im (H_+H^*_0 - H_-H^*_0)\sin \chi \right]\sin 2\theta_{D^*}
\sin 2\theta_{\rho}\Big\rbrace,
\end{array}
\label{formula:expandedDistribution}
\end{equation}
where: $\theta_{D^*}$ is the decay angle of the $D^0$ in the $D^*$ rest frame 
with respect to the $D^*$ line of flight in the $B$ rest frame; 
$\theta_{\rho}$ is the decay angle of the $\pi^-$ in the $\rho^-$ rest frame 
with respect to the $\rho^-$ line of flight in the $B$ rest frame; 
$\chi$ is the angle between the decay planes of the $D^*$ and $\rho$; 
and $\Re(x)$ and $\Im(x)$ denote the real and imaginary parts of 
$x$, respectively.  

The longitudinal and transverse polarizations are then defined as:
\begin{equation}
\displaystyle\frac{\Gamma_L}{\Gamma} =
\displaystyle\frac{|H_0|^2}{|H_0|^2 + |H_+|^2 + |H_-|^2},
\end{equation}
and
\begin{equation}
\frac{\Gamma_T}{\Gamma} = 1 - \frac{\Gamma_L}{\Gamma}, 
\end{equation}
respectively.

Previous measurements have been performed on the $D^* \rho$ system.  
Using $0.89$ fb$^{-1}$ of $\Upsilon(4S)$ data and performing an unbinned 
two-dimensional likelihood fit to the joint $(\cos\theta_{D^*}, 
\cos\theta_{\rho})$ distribution, the CLEO Collaboration measured 
$\Gamma_L/\Gamma = 0.93 \pm 0.05 \pm 0.05$~\cite{PRD50_43}.  
Later, using $3.1$ fb$^{-1}$ of $\Upsilon(4S)$ data and performing 
an unbinned maximum likelihood fit to the joint three-dimensional 
$(\cos\theta_{D^*}, \cos\theta_{\rho}, \chi)$ distribution, along with 
the invariant $B$ and $\rho$ mass distributions, 
CLEO reported a preliminary result of 
$\Gamma_L/\Gamma = 0.878\pm 0.034\pm 0.030$~\cite{CLEOCONF98-23}.  
Both results are in agreement with the theoretical prediction 
of $0.895\pm 0.019$ for $\bar{B}^0\to D^{*+}\rho^-$~\cite{PRD42_3732}.  
Testing the factorization hypothesis would benefit from further 
reduction of the experimental uncertainty.  
We report here an improved measurement using ten times 
the data of the first measurement.  
This represents the final update of the second 
analysis and uses largely the same technique, 
but with some important improvements in both the event selection 
and the treatment of acceptance.  Our results include the data used in 
the previous analyses and hence supersede them.  

The data used in this analysis were collected at the Cornell Electron Storage 
Ring (CESR) with the CLEO detector in two configurations, 
known as CLEO~II~\cite{NIMA320_66} and CLEO~II.V~\cite{NIMA418_32}. 
The data consist of an integrated luminosity of $9.1$ fb$^{-1}$ 
collected at the $\Upsilon(4S)$ resonance, corresponding to 
$9.7\times 10^{6}$ $B\bar{B}$ events, as well as $4.6$ fb$^{-1}$ 
of continuum data at energies just below the $\Upsilon(4S)$ resonance.  
The latter is used to study the backgrounds due to the non-resonant 
$e^+e^-\to q\bar{q}$ process.

In CLEO~II, the momentum measurement of charged particles is carried out 
with a tracking system consisting of a six-layer straw-tube chamber, 
a ten-layer precision drift chamber, and a 51-layer main drift chamber.  
The tracking system operates inside a $1.5$~T superconducting solenoid.  
For charged particles, the main drift chamber also provides a measurement 
of ionization energy loss ($dE/dx$), which is used for particle 
identification.  
The CLEO~II.V detector was upgraded in two main aspects, both affecting 
charged particles.  First, the straw-tube chamber was replaced with 
a three-layer double-sided silicon vertex detector; and second, 
the gas in the main drift chamber was changed from an 
argon-ethane to a helium-propane mixture.  
Photons are detected with a 7800-crystal CsI(Tl) electromagnetic 
calorimeter, which is also inside the solenoid.  
Muons are identified with proportional chambers 
placed at various depths within the steel return yoke of the magnet.  

Charged tracks with momenta greater than $250$ MeV$/c$ are required 
to come from the interaction point and be well-measured 
(based on the quality of the track fit and the number of hits).  
Identified electrons and muons are excluded, and pions 
and kaons are required to have a measured $dE/dx$ within $2.5$ standard 
deviations ($\sigma$) of their expected values.  To keep the efficiency high, 
softer tracks are only required to satisfy a looser requirement 
of consistency with originating at the interaction point.  
The $\pi^0$ candidates are formed from pairs 
of photons with an invariant mass within $2.5$ standard deviations 
of the known $\pi^0$ mass. These pairs are then kinematically fitted 
with their invariant mass constrained to the known $\pi^0$ mass.  
The $\chi^2$ of this kinematic fit must be less than nine.  
To suppress background from fake photons, the constituent photons 
of the $\pi^0$ must be detected in the central barrel calorimeter 
(which has the least material shadowing it) 
and have a minimum energy of $30-65$ MeV, 
depending on the source ($D^{*0}$, $D^0$, or $\rho^-$) of the $\pi^0$.

We reconstruct candidate $D^{*0}$ and $D^{*+}$ mesons in the modes 
$D^{*0}\to D^0\pi^0$ and $D^{*+}\to D^0\pi^+$, 
with $D^0\to K^-\pi^+$, $D^0\to K^-\pi^+\pi^0$, 
or $D^0\to K^-\pi^+\pi^-\pi^+$.  
Throughout this paper, charge conjugate modes are implied.  
The reconstructed $D^* - D$ mass differences and the $D^0$ invariant mass 
are required to be within $2.5\sigma$ of the nominal values.  
The resolutions of these quantities are obtained from Monte Carlo simulations, 
and the $D^0\to K^-\pi^+\pi^0$ resolution includes a $\pi^0$ energy 
dependence.  
We also require the $D^0\to K^-\pi^+\pi^0$ candidates to come from 
the more densely populated regions of the Dalitz plot to suppress 
combinatoric background.  
Candidate $\rho^-$ mesons are selected from $\pi^- \pi^0$ combinations 
which have an invariant mass within 150 MeV$/c^2$ of the nominal 
$\rho^-$ mass.  

The $B^-$ and $\bar{B}^0$ mesons are reconstructed by combining the 
$D^{*0}$ or $D^{*+}$ candidates with the $\rho^-$.  
We calculate a beam-constrained 
$B$ mass by substituting the beam energy ($E_b$) for the measured 
$B^-$ or $\bar{B}^0$ candidate energy ($\sum_i E_i$): 
$M \equiv \sqrt{E_b^2 - {p_B}^2}$, where $p_B$ is the 
measured momentum of the $B$ candidate.  
This improves the $M$ resolution by one order of magnitude, 
to about $3$ MeV$/c^2$.  The difference between 
the reconstructed energy of the $B^-$ or $\bar{B}^0$ candidates and 
the beam energy, $\Delta E = \sum_i E_i - E_b$, is required to be 
$0$ to within $2.5\sigma$.  
The resolution of the energy difference varies from $10$ MeV to $35$ MeV, 
depending on the decay mode, and is also obtained from Monte Carlo 
simulations.  
We also account for the dependence of our $\Delta E$ resolution 
on the $\pi^0$ (from the $\rho^-$ decay) energy, 
parameterizing it as a function of $\cos\theta_{\rho}$.  

To suppress background from the continuum under the $\Upsilon(4S)$ resonance, 
only events with a ratio of Fox-Wolfram moments~\cite{PRL41_1581}, 
$R_2 < 0.5$, are used, taking advantage of the fact that 
this ratio is larger for the more jet-like events from the continuum 
than the more spherical events from $B\bar{B}$ decays of the 
$\Upsilon(4S)$.  
We require that the polar angle of the reconstructed $B$ satisfies 
$|\cos\theta_B| < 0.95$, given the known $\sin^2\theta_B$ distribution 
for $\Upsilon(4S)$ decay.  
Finally, a requirement 
is made on the cosine of the sphericity angle, $\Theta_S$, defined as 
the angle between the sphericity axis of the $B$ decay products 
and the rest of the particles in the event.  
Because of their two-jet structure, continuum events peak strongly 
at $|\cos\Theta_S|=1$, while signal events are flat.  
Only candidates with $|\cos\Theta_S|$ below a maximum allowed value, 
dependent on the $D^0$ decay mode, are retained.  

To measure both the branching fractions and the polarization in the decay 
of $B \to D^* \rho$, we perform an unbinned maximum likelihood fit to 
extract the number of signal events and the helicity amplitudes 
from the data.  
Events from the three $D^0$ decay modes are combined in the fit.  
The likelihood function, ${\cal L}$, has the form:
\begin{equation}
  \mathcal{L} = \prod^3_{j=1}
  \displaystyle\frac{e^{-\nu_j}\nu^{n_j}_j}{n_j!}\prod^{n_j}_{i=1}
  \displaystyle\frac
  {n_j^S\cdot
   \mathcal{P}_{ji}^S(M, m, \cos\theta_{D^*}, \cos\theta_{\rho}, \chi) +
   n_j^B\cdot
   \mathcal{P}_{ji}^B(M, m, \cos\theta_{D^*}, \cos\theta_{\rho}, \chi)}
  {n_j^S + n_j^B},
\end{equation}
where $m$ is the invariant mass of the candidate $\rho^-$ meson, $n_j^S$ 
and $n_j^B$ are the number of signal and background events 
for the $j$-th $D^0$ decay mode, respectively ($\nu_j = n_j^S + n_j^B$), 
and $n_j$ is the total number of data events for the $j$-th $D^0$ decay mode.  

The signal probability distribution function, 
$\mathcal{P}_{ji}^S(M, m, \cos\theta_{D^*}, \cos\theta_{\rho}, \chi)$, 
is composed of two parts.  
The mass distribution part is a product of the beam-constrained $B$ invariant 
mass distribution (assumed to have a Gaussian probability
distribution) and a Blatt-Weisskopf form-factor-modeled~\cite{TNP1952} 
Breit-Wigner shape for the $\rho^-$ invariant mass distribution; 
the angular distribution part is given by 
Eq.~(\ref{formula:expandedDistribution}), multiplied by the detector 
acceptance, $\epsilon(\cos\theta_{D^*}, \cos\theta_{\rho}, \chi)$.

The background probability distribution function, 
$\mathcal{P}_{ji}^B(M, m, \cos\theta_{D^*}, \cos\theta_{\rho}, \chi)$, 
also has two components: the product of an ARGUS-type background 
function~\cite{PLB241_278} for the beam-constrained $B$ invariant mass and 
a flat distribution for the $\pi^-\pi^0$ invariant mass distribution; 
and an angular distribution for the background determined from events 
in the $B$ mass sideband, defined as $5.200 < M < 5.265$ GeV$/c^2$.
The background shape is parameterized as a product of second-order polynomials 
in $\cos\theta_{D^*}$ and $\cos\theta_{\rho}$ and the function 
$1 + P_{\chi}\cos(\chi + \chi_0)$ in $\chi$, where $P_{\chi}$ and $\chi_0$ 
are allowed to vary in the fit to the sideband data.  

\begin{figure}[htbp]
  \centerline{\includegraphics[width=8.6cm]{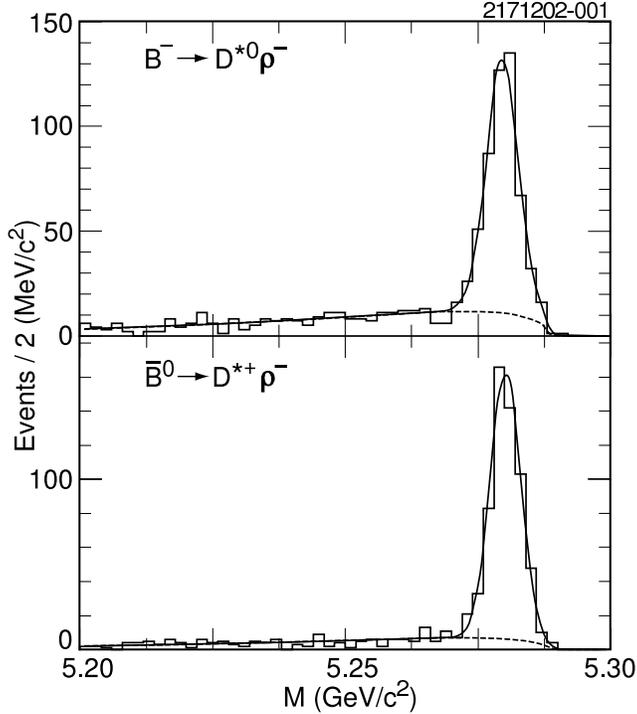}}
  \caption{$B^-$ (top) and $\bar{B}^0$ (bottom) candidate mass distributions 
           from the data, along with the results of the fits. Dashed curves 
           indicate the ARGUS-type background.}
  \label{figure:MB2}
\end{figure}

To extract the number of signal and background events, the reconstructed 
candidates with $5.20 < M < 5.30$ GeV/$c^2$ are fit, 
and the angular distributions in both the signal and background probability 
density functions are ignored.  
Figure~\ref{figure:MB2} shows the beam-constrained mass distributions 
for both $B^-$ and $\bar{B}^0$.

The efficiencies of the $B$ selection procedure for the three $D^0$ 
decay modes are determined from Monte Carlo simulations.  
Table~\ref{table:NSigEff} gives the number of signal events 
and the efficiencies, where the uncertainties are statistical only.
\begin{table}[hbt]
  \caption{Number of signal events and the efficiencies for 
           $B^- \to D^{*0}\rho^-$ and $\bar{B}^0 \to D^{*+}\rho^-$ 
           for the three $D^0$ decay modes.  
           The uncertainties shown are statistical only.}
  \begin{center}
  \begin{tabular}{|c|c|c|c|}
  \hline
  $B$ type    & $D^0$ decay mode     &         $n^S$        & $\epsilon \;\;(\%)$\\
  \hline
              & $K^-\pi^+$           & $\; 148.9\pm 13.8\;$ & $\;  6.56\pm 0.04\;$\\
  \cline{2-4}
     $B^-$    & $K^-\pi^+\pi^0$      & $\; 177.4\pm 16.6\;$ & $\;  2.20\pm 0.02\;$\\
  \cline{2-4}
              & $K^-\pi^+\pi^-\pi^+$ & $\; 136.0\pm 15.2\;$ & $\;  3.04\pm 0.03\;$\\
  \hline
              & $K^-\pi^+$           & $\; 196.3\pm 14.6\;$ & $\; 10.88\pm 0.05\;$\\
  \cline{2-4}
  $\bar{B}^0$ & $K^-\pi^+\pi^0$      & $\; 196.1\pm 16.4\;$ & $\;  3.67\pm 0.03\;$\\
  \cline{2-4}
              & $K^-\pi^+\pi^-\pi^+$ & $\; 170.6\pm 13.9\;$ & $\;  4.46\pm 0.03\;$\\
  \hline
  \end{tabular}
  \end{center}
  \label{table:NSigEff}
\end{table}

Assuming equal production of $B^+B^-$ and $B^0\bar{B}^0$ at the 
$\Upsilon(4S)$, the resulting measured branching fractions are 
${\cal B}(B^-\to D^{*0}\rho^-) = ( 0.98\;\pm\; 0.06\;\pm\; 0.16\;\pm\; 0.05)\%$ 
and 
${\cal B}(\bar{B}^0\to D^{*+}\rho^-) = ( 0.68\;\pm\; 0.03\;\pm\; 0.09\;\pm\; 0.02)\%$, 
which compare well with previous measurements~\cite{PRD50_43}, 
the recent {\sc BaBar} measurement~\cite{IJMPA16_440}, 
and the world average~\cite{PDG}.  
A statistical uncertainty and 
two systematic uncertainties are quoted in the branching fractions. 
The first systematic error includes uncertainties in the number 
of produced $B\bar{B}$ pairs ($2\%$), 
the background shape ($3\%$), our Monte Carlo statistics ($1-2\%$), 
and the charged particle tracking and $\pi^0$ detection efficiencies 
($10-18\%$). The second systematic error comes from uncertainties 
in the $D^*$ and $D^0$ decay branching fractions. The contributions 
from non-resonant $D^*\pi\pi^0$ and other non-$\rho^-$ components 
are small~\cite{PRD50_43} and neglected. 
The contribution from the helicity amplitude dependence of the efficiency 
is less than $11\%$ of the corresponding contribution from 
the Monte Carlo statistics, and hence, is also ignored.

These branching fraction measurements and the BSW prediction for 
${\cal B}(B^-\to D^{*0}\rho^-)/{\cal B}(\bar{B}^0\to D^{*+}\rho^-)$
~\cite{Neubert92}, can be used to extract 
the ratio of the effective coupling strengths for color-suppressed modes 
($a_2$) and color-enhanced modes ($a_1$) for the $D^* \rho$ final state.  
The extraction of $a_2/a_1$ is sensitive to the 
$B^+B^-$ and $B^0\bar{B}^0$ production fractions; 
we used $f_{+-}/f_{00} = 1.072 \pm 0.045 \pm 0.027 \pm 0.024$~\cite{PDG}. 
Our data give $a_2/a_1 =  0.21\pm 0.03\pm 0.05\pm 0.04\pm 0.04$, 
where the fourth uncertainty, from $f_{+-}/f_{00}$, 
is important here since other experimental systematics partially cancel.  
This result is in good agreement with the previous 
CLEO measurement~\cite{PRD50_43} and others~\cite{PPNPD35_81}.

To extract the helicity amplitudes from the data, only the reconstructed 
$B$ events in the $B$ signal region (defined as $5.27 < M <5.30$ GeV$/c^2$) 
are included in the fit.  The number of signal and background events 
for the three $D^0$ decay modes are taken from the previous mass fit, 
with the latter scaled to the $B$ signal region.

The dependence of the acceptance on the decay angles, combined with 
the effects of detector resolution, are determined from 
Monte Carlo simulations.  
Due to the {\it a priori} unknown helicity amplitudes, a weighting technique 
based on Eq.~(\ref{formula:expandedDistribution}) is employed to adjust 
the acceptance for different helicity amplitudes.  
Fits are iterated until the helicity amplitudes used 
for weighting and those resulting from the fit converge. 
Our study shows that the acceptance over the angle $\chi$ is quite flat 
and thus the acceptance can be factorized as 
$\epsilon(\cos\theta_{D^*}, \cos\theta_{\rho}, \chi) = 
 \epsilon_2(\cos\theta_{D^*}, \cos\theta_{\rho})\epsilon_1(\chi)$ 
with $\epsilon_1(\chi) = Q_0 (1 + Q_1\sin\chi  + Q_2\cos\chi +
                                  Q_3\sin2\chi + Q_4\cos2\chi)$ 
and $Q_{1,2,3,4}$ are all found to be small.  
We use the following functional form to fit the two-dimensional acceptance 
from the weighted Monte Carlo: 
\begin{equation}
\begin{array}{l}
\epsilon_2(\cos\theta_{D^*}, \cos\theta_{\rho})  =  \\
\quad\quad\quad\quad \epsilon_{00}
\displaystyle
\frac{1 + P_1\cos^2\theta_{D^*} + P_2\cos^2\theta_{\rho}
        + P_3\cos^2\theta_{D^*}\cos^2\theta_{\rho}}
     {1 + P_{10}\cos^2\theta_{D^*} + P_{11}\cos^2\theta_{\rho}
        + P_{12}\cos^2\theta_{D^*}\cos^2\theta_{\rho}}\\
\quad\quad \cdot \displaystyle \exp (-P_4\cos\theta_{D^*}
          -P_5\cos^2\theta_{D^*}
          -P_6\cos^3\theta_{D^*}
          -P_7\cos\theta_{\rho}
          -P_8\cos^2\theta_{\rho}
          -P_9\cos^3\theta_{\rho}).  
\end{array}
\end{equation}
This gives an excellent fit with no discernible pattern of residuals.  

\begin{figure*}[htb]
  \centerline{\includegraphics[width=\textwidth]{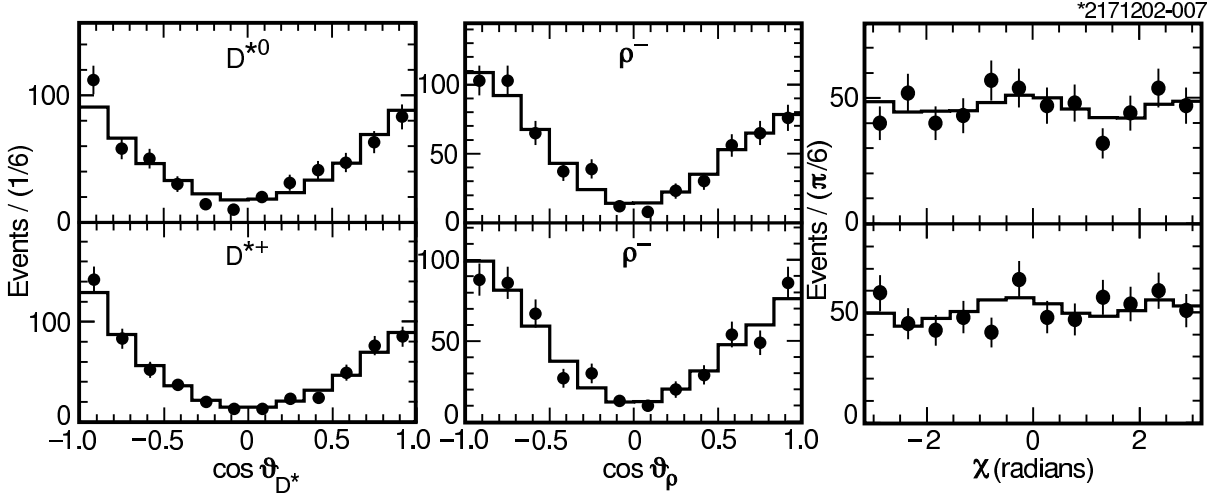}}
  \caption{The $\cos\theta_{D^*}$ (left), 
               $\cos\theta_{\rho}$ (middle)
           and $\chi$ (right) distributions 
               for $B^-\to D^{*0}\rho^-$ (top) 
               and $\bar{B}^0\to D^{*+}\rho^-$ (bottom) 
               from the data (dots) 
               and the corresponding fit projections (histograms). }
  \label{figure:AB}
\end{figure*}

Performing the unbinned maximum likelihood fit, we obtain the helicity 
amplitudes for the decay $B\to D^*\rho$ listed in
 Table~\ref{table:HelAmpSummary}.  
Figure~\ref{figure:AB} shows the one-dimensional 
angular distributions and the projections from the fit.  
The errors quoted in the table are the statistical and 
systematic uncertainties, respectively.  

\begin{table}[htbp]
  \caption{The measured helicity amplitudes 
           for $B^-\to D^{*0}\rho^-$ and $\bar{B}^0\to D^{*+}\rho^-$.  
           The phase of $H_0$ is fixed to zero in each mode.
           $\alpha_+$ and $\alpha_-$ are the phases, in radians, 
           of $H_+$ and $H_-$, respectively: 
           $H_\pm = |H_\pm| \exp (i \alpha_\pm)$.}
  \begin{center}
  \begin{tabular}{|c|c|c|}
  \hline
             &                                 & \\[-4mm]
  Quantity   & $B^-\to D^{*0}\rho^-$           & $\bar{B}^0\to D^{*+}\rho^-$\\
  \hline
  $|H_0|$    & $\;0.944 \pm 0.009 \pm 0.009\;$ & $\;0.941 \pm 0.009 \pm 0.006\;$\\
  \hline
  $|H_+|$    & $\;0.122 \pm 0.040 \pm 0.010\;$ & $\;0.107 \pm 0.031 \pm 0.011\;$\\
  \hline
  $\alpha_+$ & $\;1.02 \pm 0.28 \pm 0.11\;$    & $\;1.42 \pm 0.27 \pm 0.04\;$\\
  \hline
  $|H_-|$    & $\;0.306 \pm 0.030 \pm 0.025\;$ & $\;0.322 \pm 0.025 \pm 0.016\;$\\
  \hline
  $\alpha_-$ & $\;0.65 \pm 0.16 \pm 0.04\;$    & $\;0.31 \pm 0.12 \pm 0.04\;$\\
  \hline
  \end{tabular}
  \end{center}
  \label{table:HelAmpSummary}
\end{table}

The sources of systematic uncertainty are the acceptance parameterization, 
detector smearing, background level and shape, non-resonant $\pi^-\pi^0$ 
contribution, and the polarization dependence on the mass 
of the $\rho^-$ meson.  
Their individual contributions are shown in Table~\ref{table:SummarySysErr}.  
For the acceptance parameterization, we use different functional forms 
for both $\epsilon_2(\cos\theta_{D^*}, \cos\theta_{\rho})$ and 
$\epsilon_1(\chi)$ in the maximum likelihood fit, and the changes 
of the helicity amplitudes are taken as the systematic uncertainties.  
To gauge the effect of detector smearing, we increase the smearing 
in the nominal Monte Carlo by a conservative 15\% of itself.  
We increase the number of background events for each $D^0$ decay mode 
independently by $1\sigma$ and use the observed shifts as the corresponding 
systematic uncertainties.  
Our largest systematic uncertainty comes from the shape of the 
background angular distribution.  
We compare the nominal fit results to three other fits: 
one with the background flat in the decay angles, one with the shape fit 
to the Monte Carlo events in the $M$ sideband, and one with the shape fit 
to the non-signal Monte Carlo events in the $M$ peak region.  
The largest variation among these three is taken as the systematic 
uncertainty. 
To account for a background contribution from non-resonant $\pi^-\pi^0$ 
combinations in the data sample, an $S-$wave component with 
a flat angular distribution is added in the helicity angular distribution.  
Finally, we include a systematic uncertainty due to our sensitivity 
to the $q^2$ dependence of the helicity amplitudes.  
In the nominal fit, we ignore this possible dependence.  
Instead, we relate the helicity amplitudes at a momentum transfer 
$q^2 = M_{\rho}^2$ to the actual $q^2$ of the events, using 
the factorization hypothesis, and we use the shifts as the uncertainty.  
The total systematic uncertainty is then the sum of the above contributions 
in quadrature.  
\begin{table*}[htbp]
  \caption{Summary of the systematic uncertainties for 
           the helicity amplitudes of $B^-\to D^{*0}\rho^-$ 
           and $\bar{B}^0\to D^{*+}\rho^-$, respectively.}
  \begin{center}
  \begin{tabular}{|c|c|c|c|c|c|c|c|}
  \hline
  $D^{*0}\rho^-$    & \multicolumn{7}{c|}{All $\times 10^{-2}$}\\
  \hline
  Quantity & Accep. & Smearing & Bkg. level & Bkg. shape & Non-res. & $q^2$ dep. & Total\\
  \hline
  $|H_0|$           & 0.08 & 0.06 & 0.13 & 0.83 & 0.01 & 0.07 & 0.85\\
  \hline
  $|H_+|$           & 0.32 & 0.25 & 0.37 & 0.60 & 0.23 & 0.43 & 0.95\\
  \hline
  $\alpha_+$        & 2.23 & 1.90 & 2.96 & 9.82 & 0.71 & 1.59 & 10.81\\
  \hline
  $|H_-|$           & 0.24 & 0.28 & 0.30 & 2.45 & 0.08 & 0.03 & 2.50\\
  \hline
  $\alpha_-$        & 1.29 & 1.12 & 2.22 & 1.07 & 1.02 & 2.84 & 4.25\\
  \hline
  $\Gamma_L/\Gamma$ & 0.14 & 0.11 & 0.25 & 1.58 & 0.01 & 0.12 & 1.62\\
  \hline\hline
  $D^{*+}\rho^-$    & \multicolumn{7}{c|}{All $\times 10^{-2}$}\\
  \hline
  Quantity & Accep. & Smearing & Bkg. level & Bkg. shape & Non-res. & $q^2$ dep. & Total\\
  \hline
  $|H_0|$           & 0.07 & 0.01 & 0.04 & 0.61 & 0.13 & 0.10 & 0.63\\
  \hline
  $|H_+|$           & 0.25 & 0.01 & 0.09 & 0.92 & 0.44 & 0.09 & 1.06\\
  \hline
  $\alpha_+$        & 2.17 & 0.18 & 1.64 & 2.12 & 2.17 & 1.35 & 4.29\\
  \hline
  $|H_-|$           & 0.20 & 0.03 & 0.13 & 1.52 & 0.06 & 0.26 & 1.56\\
  \hline
  $\alpha_-$        & 0.92 & 0.01 & 0.39 & 3.95 & 0.18 & 0.16 & 4.08\\
  \hline
  $\Gamma_L/\Gamma$ & 0.13 & 0.02 & 0.07 & 1.14 & 0.25 & 0.19 & 1.19\\
  \hline
  \end{tabular}
  \end{center}
  \label{table:SummarySysErr}
\end{table*}

As can be seen from Table~\ref{table:HelAmpSummary}, our results 
indicate possible non-trivial helicity amplitude phases, $\alpha_+$ and 
$\alpha_-$.    
To better gauge the significance of such a conclusion, we use the quantity 
$\sqrt{\Delta (-2\ln{\mathcal{L}_{max}})}$, where $\Delta$ 
refers to the increase in $-2\ln{\mathcal{L}_{max}}$ when both 
phases are forced to be zero, as compared to the nominal fit with 
floating phases.  Interpreting this quantity as the net statistical 
significance of non-zero phases, we find 
$ 3.19\sigma$ and $ 2.75\sigma$ 
for $B^-\to D^{*0}\rho^-$ and $\bar{B}^0\to D^{*+}\rho^-$, respectively.  
The stability of the significance is evaluated by examining the changes 
in $\sqrt{\Delta (-2\ln{\mathcal{L}_{max}})}$ for all of the 
systematic variations discussed above.  
The values are quite stable, ranging from $3.09-3.54\sigma$ 
and $2.65-2.83\sigma$ for $B^-$ and $\bar{B}^0$, 
respectively.  
Previously, indications of FSI phases have also been reported 
in the $D\pi$~\cite{PRL88_062001} 
and the $J/\psi K^*$ systems~\cite{PRL85_4668}.

\begin{figure}[htb]
  \centerline{\includegraphics[width=8.6cm]{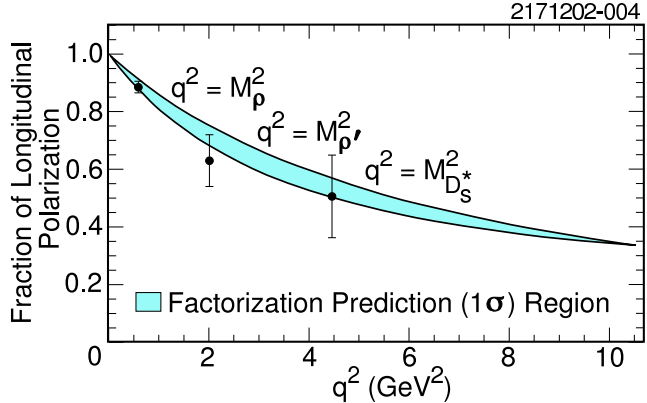}}
  \caption{The fraction of longitudinal polarization in 
           $\bar{B}^0\to D^{*+}X$ decays as a function of  
           $q^2 = M_X^2$, where $X$ is a vector meson.  
           Shown are the current $\bar{B}^0\to D^{*+}\rho^-$ polarization 
           measurement, and earlier measurements 
           of $\bar{B}^0\to D^{*+}\rho^{\prime -}$~\cite{PRD64_092001},
           and $\bar{B}^0\to D^{*+}D_s^{*-}$~\cite{PRD62_112003}.  
           The shaded region represents the prediction using factorization 
           and HQET, and extrapolating from the semileptonic 
           $\bar{B}^0\to D^{*+}\ell^-\bar{\nu}$ form factor 
           results~\cite{PRL76_3898}.  
           The shaded contour shows a one standard deviation variation 
           in the theoretical prediction.}
  \label{figure:FracL}
\end{figure}

The results for the helicity amplitudes correspond to 
\begin{equation}
  \left\{
  \begin{array}{lcl}
  \displaystyle\frac{\Gamma_L}{\Gamma}(B^-\to D^{*0}\rho^-)
     & = & 0.892 \pm 0.018 \pm 0.016,\\[4mm]
  \displaystyle\frac{\Gamma_L}{\Gamma}(\bar{B}^0\to D^{*+}\rho^-)
     & = & 0.885 \pm 0.016 \pm 0.012,
  \end{array}
  \right.
\end{equation}
where the two uncertainties are statistical and systematic, respectively.  
Within the uncertainties, the fraction of longitudinal polarization for 
$\bar{B}^0\to D^{*+}\rho^-$ is in good agreement with the previous 
CLEO measurement~\cite{PRD50_43} and with the HQET prediction of 
$0.895\pm 0.019$~\cite{PRD42_3732} using factorization 
and the measurements of the semileptonic form factors.  
Longitudinal polarization as a function of $q^2$ is plotted in 
Figure~\ref{figure:FracL} for such a prediction and compared with our 
new $D^{*+}\rho^-$ result, as well as 
previous measurements for $D^{*+}\rho^{\prime -}$~\cite{PRD64_092001} and
$D^{*+}D_s^{*-}$~\cite{PRD62_112003}.  
The agreement is excellent, indicating that the factorization hypothesis 
works well at the level of the current uncertainties.  

In summary, we have measured both the branching fractions 
and the helicity amplitudes for $B \to D^* \rho$.  
The values of the branching fractions, the ratio $a_2/a_1$, 
and the degree of longitudinal polarization are in good agreement 
with previous measurements and with theoretical predictions.  
The measurement of the fraction of longitudinal polarization confirms 
the validity of the factorization assumption at relatively low $q^2$. 
Finally, the measurement of the helicity amplitudes indicates a strong 
possibility of non-trivial helicity amplitude phases which would arise 
from final-state interactions. 
Such phases are of interest since they are required 
for the observation of direct $CP$ violation in $B$ 
decay rates~\cite{EPJC6_647}.  

We gratefully acknowledge the effort of the CESR staff 
in providing us with
excellent luminosity and running conditions.
We thank Lalith Perera for contributions to a 
preliminary version of this analysis. 
M. Selen thanks the Research Corporation, 
and A.H. Mahmood thanks the Texas Advanced Research Program.
This work was supported by the 
National Science Foundation, 
and the
U.S. Department of Energy.



\begin{thebibliography}{99}
\bibitem{ZPC34_103} M. Bauer, B. Stech and M. Wirbel, 
         Z. Phys. C {\bf 34}, 103 (1987).
\bibitem{PL89B_105} J. K\"{o}rner and G. Goldstein, 
         Phys. Lett. {\bf 89 B}, 105 (1979).
\bibitem{PRD50_43} CLEO Collaboration, 
         M.S. Alam {\it et al.}, Phys. Rev. {\bf D 50}, 43 (1994).
\bibitem{CLEOCONF98-23} CLEO Collaboration, G. Bonvicini {\it et al.}, 
         CLEO CONF 98-23, International Conference on High-Energy Physics 
         (Vancouver, 1998).
\bibitem{PRD42_3732} 
         J.L. Rosner, Phys. Rev. {\bf D 42}, 3732 (1990); 
         M. Neubert, Phys. Lett. {\bf B 264}, 455 (1991); 
         G. Kramer, T. Mannel and W.F. Palmer, 
                    Z. Phys. {\bf C 55}, 497 (1992); 
         J.D. Richman, in 
              {\it Probing the Standard Model of Particle Interactions}, 
              edited by R. Gupta, A. Morel, E. de Rafael, and F. David 
              (Elsevier, Amsterdam, 1999), p. 640.
\bibitem{NIMA320_66} CLEO Collaboration, Y. Kubota {\it et al.}, 
         Nucl. Instrum. Methods Phys. Res., Sect. {\bf A 320}, 66 (1992).
\bibitem{NIMA418_32} T.S. Hill, 
         Nucl. Instrum. Methods Phys. Res., Sect. {\bf A 418}, 32 (1998).
\bibitem{PRL41_1581} G. Fox and S. Wolfram,
         Phys. Rev. Lett. {\bf 41}, 1581 (1978).
\bibitem{TNP1952} V.F. Weisskopf and J.M. Blatt, 
         {\it Theoretical Nuclear Physics}, p. 361 (Wiley, 1952).
\bibitem{PLB241_278} ARGUS Collaboration, H. Albrecht {\it et al}., 
         Phys. Lett. {\bf B 241}, 278 (1990); {\bf B 254}, 288 (1991).
\bibitem{IJMPA16_440} B. Brau,
         Int. J. Mod. Phys. {\bf A 16}, Suppl. 1A 440 (2001).
\bibitem{PDG} Particle Data Group, Phys. Rev. D {\bf 66}, 010001 (2002).
\bibitem{Neubert92} M. Neubert, V. Rieckert, Q.P. Xu and B. Stech, 
         in {\it Heavy Flavours}, edited by A.J. Buras and H. Lindner 
         (World Scientific, Singapore, 1992).
\bibitem{PPNPD35_81} T.E. Browder and K. Honscheid, 
         Prog. Part. Nucl. Phys. {\bf 35}, 81 (1995); 
         Particle Data Group, Eur. Phys. J. {\bf C 15}, 1 (2000).
\bibitem{PRL88_062001} CLEO Collaboration, T.E. Coan {\it et al}., 
         Phys. Rev. Lett. {\bf 88}, 062001 (2002); 
         CLEO Collaboration, S. Ahmed {\it et al}., 
         Phys. Rev. {\bf D66}, 031101 (2002).
\bibitem{PRL85_4668} CDF Collaboration, T. Affolder {\it et al}., 
         Phys. Rev. Lett. {\bf 85}, 4668 (2000).
\bibitem{PRD64_092001} CLEO Collaboration, J.P. Alexander {\it et al}.,
         Phys. Rev. {\bf D 64}, 092001 (2001).
\bibitem{PRD62_112003} CLEO Collaboration, S. Ahmed {\it et al}.,
         Phys. Rev. {\bf D 62}, 112003 (2000).
\bibitem{PRL76_3898} CLEO Collaboration, J. Duboscq {\it et al.}, 
         Phys. Rev. Lett. {\bf 76}, 3898 (1996).
\bibitem{EPJC6_647} A.S. Dighe, I. Dunietz and R. Fleischer,
         Eur. Phys. J. {\bf C 6}, 647 (1999).
\end{thebibliography}
\end{document}